\documentstyle[12pt]{article}

\catcode`\@=11
\long\def\@makefntext#1{
\protect\noindent \hbox to 3.2pt {\hskip-.9pt
$^{{\ninerm\@thefnmark}}$\hfil}#1\hfill}		

\def\@makefnmark{\hbox to 0pt{$^{\@thefnmark}$\hss}}  

\def\ps@myheadings{\let\@mkboth\@gobbletwo
\def\@oddhead{\hbox{}
\rightmark\hfil\ninerm\thepage}
\def\@oddfoot{}\def\@evenhead{\ninerm\thepage\hfil
\leftmark\hbox{}}\def\@evenfoot{}
\def\sectionmark##1{}\def\subsectionmark##1{}}

\renewcommand{\thefootnote}{\fnsymbol{footnote}}

\newcounter{sectionc}\newcounter{subsectionc}\newcounter{subsubsectionc}
\renewcommand{\section}[1] {\vspace*{0.6cm}\addtocounter{sectionc}{1}
\setcounter{subsectionc}{0}\setcounter{subsubsectionc}{0}\noindent
	{\normalsize\bf\thesectionc. #1}\par\vspace*{0.4cm}}
\renewcommand{\subsection}[1] {\vspace*{0.6cm}\addtocounter{subsectionc}{1}
	\setcounter{subsubsectionc}{0}\noindent
	{\normalsize\it\thesectionc.\thesubsectionc. #1}\par\vspace*{0.4cm}}
\renewcommand{\subsubsection}[1]
{\vspace*{0.6cm}\addtocounter{subsubsectionc}{1}
	\noindent {\normalsize\rm\thesectionc.\thesubsectionc.\thesubsubsectionc.
	#1}\par\vspace*{0.4cm}}

\newcounter{appendixc}
\newcounter{subappendixc}[appendixc]
\newcounter{subsubappendixc}[subappendixc]

\renewcommand{\appendix}[1] {\vspace*{0.6cm}
        \refstepcounter{appendixc}
        \setcounter{figure}{0}
        \setcounter{table}{0}
        \setcounter{equation}{0}
        \renewcommand{\thefigure}{\Alph{appendixc}.\arabic{figure}}
        \renewcommand{\thetable}{\Alph{appendixc}.\arabic{table}}
        \renewcommand{\theappendixc}{\Alph{appendixc}}
        \renewcommand{\theequation}{\Alph{appendixc}.\arabic{equation}}
        \noindent{\bf Appendix \theappendixc #1}\par\vspace*{0.4cm}}

\def\abstracts#1{{

\centering{\begin{minipage}{12.2truecm}\footnotesize\baselineskip=12pt\noindent
	\centerline{\footnotesize ABSTRACT}\vspace*{0.3cm}
	\parindent=0pt #1
	\end{minipage}}\par}}

\renewenvironment{thebibliography}[1]
	{\begin{list}{\arabic{enumi}.}
	{\usecounter{enumi}\setlength{\parsep}{0pt}
\setlength{\leftmargin 1.25cm}{\rightmargin 0pt}
	 \setlength{\itemsep}{0pt} \settowidth
	{\labelwidth}{#1.}\sloppy}}{\end{list}}

\newcounter{itemlistc}
\newcounter{romanlistc}
\newcounter{alphlistc}
\newcounter{arabiclistc}

\newcommand{\fcaption}[1]{
        \refstepcounter{figure}
        \setbox\@tempboxa = \hbox{\footnotesize Fig.~\thefigure. #1}
        \ifdim \wd\@tempboxa > 6in
           {\begin{center}
        \parbox{6in}{\footnotesize\baselineskip=12pt Fig.~\thefigure. #1}
            \end{center}}
        \else
             {\begin{center}
             {\footnotesize Fig.~\thefigure. #1}
              \end{center}}
        \fi}

\newcommand{\tcaption}[1]{
        \refstepcounter{table}
        \setbox\@tempboxa = \hbox{\footnotesize Table~\thetable. #1}
        \ifdim \wd\@tempboxa > 6in
           {\begin{center}
        \parbox{6in}{\footnotesize\baselineskip=12pt Table~\thetable. #1}
            \end{center}}
        \else
             {\begin{center}
             {\footnotesize Table~\thetable. #1}
              \end{center}}
        \fi}

\def\@citex[#1]#2{\if@filesw\immediate\write\@auxout
	{\string\citation{#2}}\fi
\def\@citea{}\@cite{\@for\@citeb:=#2\do
	{\@citea\def\@citea{,}\@ifundefined
	{b@\@citeb}{{\bf ?}\@warning
	{Citation `\@citeb' on page \thepage \space undefined}}
	{\csname b@\@citeb\endcsname}}}{#1}}

\newif\if@cghi
\def\cite{\@cghitrue\@ifnextchar [{\@tempswatrue
	\@citex}{\@tempswafalse\@citex[]}}
\def\citelow{\@cghifalse\@ifnextchar [{\@tempswatrue
	\@citex}{\@tempswafalse\@citex[]}}
\def\@cite#1#2{{$\null^{#1}$\if@tempswa\typeout
	{IJCGA warning: optional citation argument
	ignored: `#2'} \fi}}

 1
 1
 1

\font\ninerm=cmr9

\def\tg{\tilde g}

\def\tl{\tilde \ell}

\def\tnu{\tilde \nu}

\def\tz{\widetilde Z}
\def\tw{\widetilde W}
\input{psfig}

\begin{document}

\newcommand{\st}{\scriptstyle}
\newcommand{\sst}{\scriptscriptstyle}
\newcommand{\mco}{\multicolumn}
\newcommand{\epp}{\epsilon^{\prime}}
\newcommand{\vep}{\varepsilon}
\newcommand{\ra}{\rightarrow}
\newcommand{\ppg}{\pi^+\pi^-\gamma}
\newcommand{\vp}{{\bf p}}
\newcommand{\ko}{K^0}
\newcommand{\kb}{\bar{K^0}}
\newcommand{\al}{\alpha}
\newcommand{\ab}{\bar{\alpha}}
\def\be{\begin{equation}}
\def\ee{\end{equation}}
\def\bea{\begin{eqnarray}}
\def\eea{\end{eqnarray}}
\def\CPbar{\hbox{{\rm CP}\hskip-1.80em{/}}}

\centerline{\normalsize\bf MULTI-CHANNEL SEARCH FOR SUPERGRAVITY}
\baselineskip=22pt
\centerline{\normalsize\bf AT THE}
\baselineskip=16pt
\centerline{\normalsize\bf LARGE HADRON COLLIDER}

\centerline{\footnotesize Chih-Hao Chen}
\baselineskip=13pt
\centerline{\footnotesize\it Department of Physics, Florida State University}
\baselineskip=12pt
\centerline{\footnotesize\it Tallahassee, FL 30306, USA}
\centerline{\footnotesize E-mail: chih\_hao@hep.fsu.edu}

\vspace*{0.9cm}
\abstracts{ The potential of seeing supersymmetry (SUSY) at the CERN
Large Hadron Collider (LHC) was studied
by looking at 3 types of signals: dilepton events from slepton pair
productions,
trilepton events from chargino/neutralino productions and missing energy
plus multi-jet events from gluino/squark productions. I
described my results by mapping out reachable areas in the supergravity
parameter space.  Areas explorable at LEP II were also mapped out for
comparison.}

\normalsize\baselineskip=15pt
\setcounter{footnote}{0}
\renewcommand{\thefootnote}{\alph{footnote}}
\section{Introduction}
In the supergravity model (SUGRA), gauge couplings,
masses of scalar fields, masses of gauginos,
trilinear and bilinear soft SUSY breaking terms
are assumed to unify at $M_X\sim 10^{16}$ GeV
and leaves us with only five free parameters (along with $m_t$):
\begin{displaymath}
m_0, m_\frac{1}{2}, A_0, tan\beta \mbox{ and } sgn(\mu).
\end{displaymath}
With one set of these parameters given, by evolving $26$ renormalization
group equations (RGEs)
and requiring radiative electroweak symmetry breaking,
particle/sparticle masses and couplings can be obtained
and detailed phenomenological study can be performed.

Most of this study was done in the framework of SUGRA and the tool is
ISAJET which is a Monte Carlo simulator for $pp, p\overline{p}$ and $e^+e^-$
colliders. In the SUGRA parameter space, areas which are excluded
by theories (no radiative electroweak symmetry breaking, lightest
SUSY particle (LSP) not $\tz_1$ or
tachyonic particle mass) or by experiments (
$m_{\tw_1}<47$ GeV,
$m_{\tl}<45$ GeV,
$m_{\tnu}<43$ GeV,
$m_{H_\ell}<60$ GeV and
the area excluded by non-observation of an excess of $E\llap/_T$ events from
gluino squark production at the Tevatron) were avoided.

\section{Phenomenological overview}

\subsection{Dilepton events from slepton pair productions$^1$}
It has been shown that the main sources for this type of events are
$\tl_R \overline{\tl}_R$ and $\tl_L \overline{\tl}_L$ productions
and that they should be detectable up to a limit of
$m_{\tl} \sim 200-250$ GeV.

\subsection{Trilepton events from chargino/neutralino productions$^2$}
This signature comes mainly from $\tw_1 \tz_2$ production and
can be seen if the spoiler modes (
$\tz_2 \to \tz_1 H_\ell$ or $\tz_2 \to \tz_1 Z$)
are not open and the interference effect in the
$\tz_2 \to \tz_1 \ell \overline \ell$ branching ratio
(which is especially noticeable when $\mu>0$)
is small. Special cuts were developed to suppress $t\overline t$
background which could otherwise be severe.

\subsection{$E\llap/_T$ plus multi-jet events from gluino/squark
productions$^3$}
This is the most powerful channel to see SUSY at LHC because it reaches
out to a huge area in the parameter space. $m_{\tg}$ can be measured
to $15-25\%$ by looking at hemisphere masses. Jet multiplicity may
reveal informations of mass difference between the gluino and the squark.

\section{Conclusions}
Results of this calculation is shown in fig. 1 where I have taken
$A_0=0$, $tan\beta=2$ and $\mu<0$.
The brick area is excluded by theory and the slashed area is
excluded by experiments. In fig. 1a, the contour labeled by $\tl_R(200)$
is about where we can reach in the dilepton channel. With a luminosity
of $10 fb^{-1}$, the reachable
($5\sigma$ is our criteria) area of trilepton
channel lies below the 2 spoilers when 2-body decay modes (
$\tz_2 \to \tl \ell$ and $\tz_2 \to \tnu\nu$) are closed.
When 2-body decay modes are open, we can see this signature over the spoilers
but not higher than $m_\frac{1}{2}=400$ GeV. At LEP II
the areas below $H_\ell(90)$ and $\tw_1(90)$ are explorable.
The $E\llap/_T(5\sigma)$ contour is spectacular in that it covers most of
this parameter space. In fig. 1b mass contours of gluino and squark are
plotted for comparison with 1a.

\section{Acknowledgements}
I thank Howard Baer, Frank Paige and Xerxes Tata for collaborations on these
projects. In addition I thank J. Gunion, T. Han and all UC-Davis workers for
organizing this very interesting conference.

\begin{figure}[tbh]
\centerline{\psfig{file=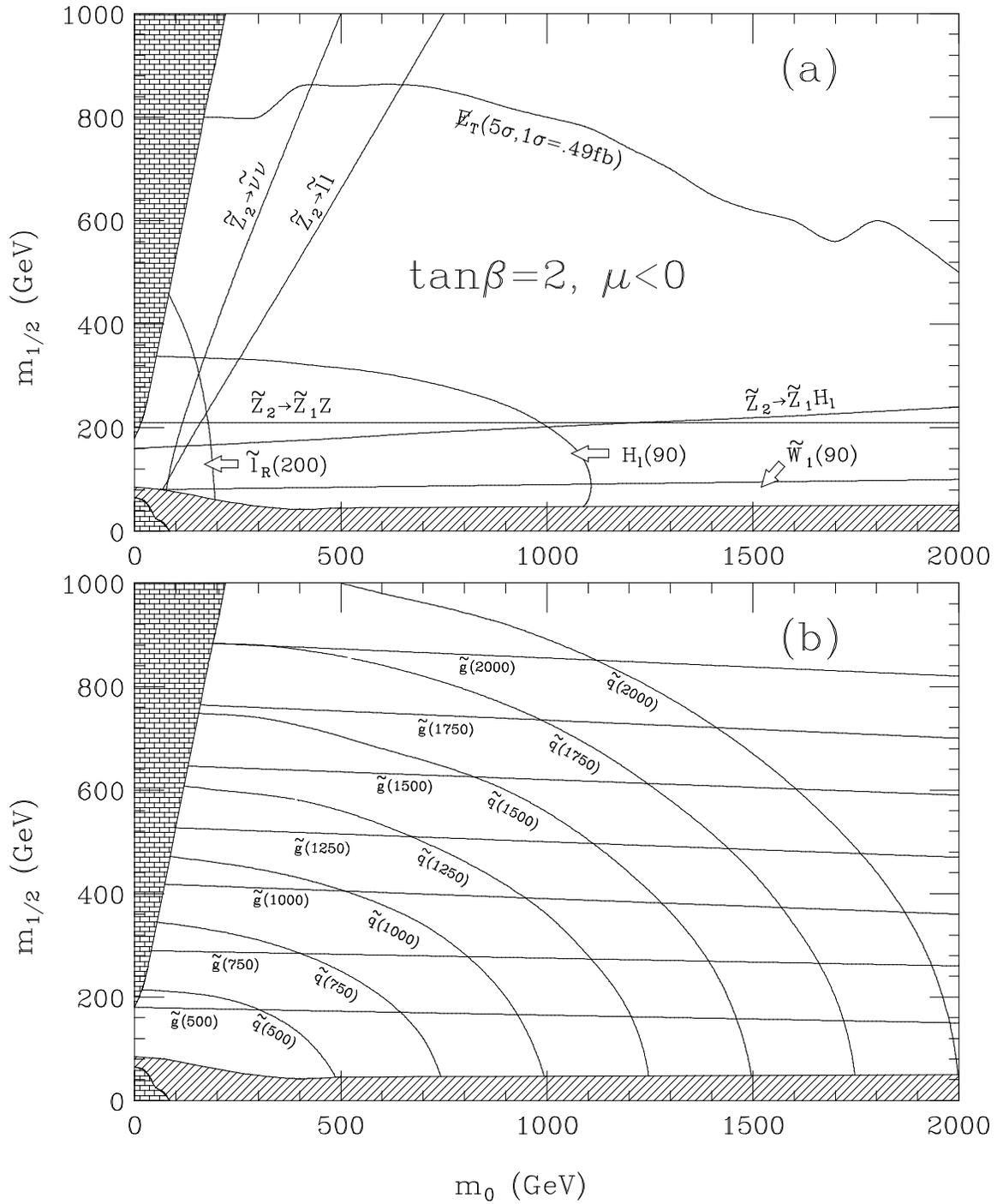,height=20cm,angle=90}}
\caption[]{Explorable regions of various channels on the $m_\frac{1}{2}$
vs. $m_0$ plane. Shaded areas are excluded by theory or experiments.
}
\label{fig1}
\end{figure}


\section{References}

\begin{thebibliography}{9}

\bibitem{one} H. Baer, C-H. Chen, F. Paige and X. Tata,
Phys. Rev. {\bf D49}, 3283 (1994).
\bibitem{two} H. Baer, C-H. Chen, F. Paige and X. Tata,
Phys. Rev. {\bf D50}, 4508 (1994).
\bibitem{three} H. Baer, C-H. Chen, F. Paige and X. Tata,
FSU-HEP-950204 (1995).

\end{thebibliography}
\end{document}